\documentclass[prl,floatfix,twocolumn,preprintnumbers,amsmath,amssymb,superscriptaddress]{revtex4}
\usepackage{color}

\usepackage{graphicx,psfrag}
\usepackage{hyperref}
\usepackage{verbatim}
\usepackage[utf8]{inputenc}

\begin{document}
\title{The de Sitter swampland conjecture and supersymmetric AdS vacua }
\author{Joseph P. Conlon}
\email{joseph.conlon@physics.ox.ac.uk}
\affiliation{Rudolf Peierls Centre for Theoretical Physics, University of Oxford,
1 Keble Rd., Oxford OX1 3NP, UK}
\begin{abstract}
It has recently been conjectured that string theory does not admit de Sitter critical points. This note points out that
in several cases, including KKLT or racetrack models, this statement is equivalent to the absence of supersymmetric
Minkowski or AdS solutions. This equivalence arises from establishing the positivity of the potential in a large-radius limit, requiring a turnover of the potential before reaching an AdS vacuum. For example, this conjecture is incompatible with the simplest 1-modulus KKLT AdS supersymmetric solution.
\end{abstract}
\maketitle

In making contact between string theory and observable physics, any general statements that apply to all string vacua are clearly of great interest. In this respect it has recently been conjectured that the effective theory of string compactifications
obeys a constraint \cite{oogurivafa} (see \cite{170903554, dvr} for related earlier work)
\begin{equation}
\label{claim}
\vert \nabla V \vert \geq c V
\end{equation}
for some non-zero positive constant $c$. Such a constraint implies the absence of any de Sitter critical points, and as such would have enormous observational consequences - if correct.

\subsection{Relationship to AdS Vacua}

Given that extraordinary claims require extraordinary evidence, it is, however, worth first investigating whether this conjecture is likely to be true (other recent work on this includes \cite{180610999, 180705193, 180706581, 180709538, 180709698, 180709794, 180803397}). The purpose of this short note is to emphasise that this conjecture - already strong - is far stronger even than it first appears. If it holds, it also rules out AdS vacua, either supersymmetric or non-supersymmetric, in what are some of the most well studied scenarios of moduli stabilisation. In particular, the existence of a 1-modulus supersymmetric KKLT AdS solution (i.e. the first step of KKLT \emph{not} including
the $\bar{D}3$-brane) would be sufficient to falsify the conjecture of Eq. (\ref{claim}).

This point is best illustrated by example. First, consider type IIB flux compactifications with a single K\"ahler modulus (various Calabi-Yaus exist with only a single K\"ahler modulus, so this is perfectly feasible). As is well-known, the resulting leading order effective no-scale supergravity theory is \cite{9908088, gkp}
\begin{eqnarray}
\label{noscale}
K & = & - 3 \ln (T + \bar{T}), \nonumber \\
W & = & W_0.
\end{eqnarray}
The large multiplicity of fluxes makes it possible to choose fluxes so that $W_0$ is very small \cite{denefdouglas}. In this case, the dilaton and complex structure moduli are stabilised at large masses ($m_U, m_S \gg m_{3/2}$) far above the scale of the dynamics of the K\"ahler moduli.
It is also possible to chose the 3-form fluxes so as to saturate the D3-brane tadpole, and thereby avoid the presence of any wandering D3 branes (and associated moduli) in the theory. This makes a strong case that there exist compactifications for which the effective low-energy dynamics can be described at leading order by the 1-modulus no-scale theory of Eq. (\ref{noscale}).

Now consider going beyond the no-scale limit. In the limit of ${\rm Re} (T) \to \infty$, the potential is dominated by perturbative corrections to $K$ that lift the no-scale structure. The sign of the potential at large volume is set by the leading correction as $T \to \infty$.
As is well known, the dominant correction arises from the $\mathcal{R}^4$ $(\alpha')^{3}$ correction \cite{bbhl, cqs}, and in the case where $h^{1,1} < h^{2,1}$ this results in a positive contribution to the scalar potential. Although there exist $(\alpha')^{2}$ corrections to the K\"ahler potential \cite{bhk}, at the level of the scalar potential these correspond (due to extended no-scale structure \cite{CCQ}) to $(\alpha')^{4}$ corrections to the scalar potential that are subleading at large volume to the $\mathcal{R}^4$ correction \cite{BergHaackKors, CCQ}). As the $\mathcal{R}^4$ correction then represents the dominant correction as ${\rm Re} (T) \to \infty$, this establishes that the potential approaches zero from above as $\hbox{\rm T} \to \infty$.

Now suppose that quantum corrections cause this system to have a supersymmetric vacuum (for example as in KKLT with a non-perturbative superpotential $W_0 + A e^{-a T}$, but the argument holds for any such solution - although note the argument in \cite{0405011} against such instanton corrections in 1-modulus models). As any supersymmetric vacuum necessarily has either zero or negative vanishing energy, it follows that for such solutions to exist the potential must turn over coming in from large $T$ to the location of the supersymmetric solution. Even without knowledge of the precise form or location of the supersymmetric solution, this then implies (by continuity) the existence of a de Sitter critical point at some point in moduli space.

Note both that, so long as the theory is described by a quantum-corrected version of Eq. \ref{noscale}, this argument holds even if the supersymmetic solution is located deep in an uncalculable strong coupling region of moduli space, and also that the argument is unaffected by $T \equiv \tau + i a$ containing two scalar fields, as the imaginary part $a$ is an axionic field and so has a compact and periodic moduli space. In particular, the compactness of the axionic moduli space implies that it is always possible to locate a critical point in the axionic direction, reducing the problem to that of a single scalar field.

For such models, it therefore follows that the presence of a supersymmetric AdS solution implies the presence of a de Sitter critical point. This renders the conjecture of Eq. (\ref{claim}) even stronger, as it becomes not merely a statement about de Sitter - which one might regard as something exotic - but also something that can be falsified by the presence of supersymmetric AdS vacua.

A couple of comments are in order here. First, if the effective theory is described by many moduli - and in particular by many non-compact moduli - then the combination of positivity of the potential at large volume and a supersymmetric solution in the interior of moduli space does not automatically guarantee the existence of a de Sitter critical point (although it is also true that the arguments given in \cite{oogurivafa} to support the conjecture of Eq. (\ref{claim}) are based only on the overall volume modulus).
However, an appealing aspect of type IIB flux models is that there does exist a clear logic that the theory can be reduced to an effective supergravity theory described only by a single modulus, with all other moduli stabilised at a much higher scale.

Second, it is worth noting that a similar feature - guaranteed positivity of the potential in the large modulus regime -
also applies to the much studied case of dilaton stabilisation in $\mathcal{N}=1$ compactifications of the heterotic string.
Here the weak coupling behaviour is described by
\begin{eqnarray}
K & = & - \ln (S + \bar{S}), \nonumber \\
W & = & A e^{-a S} + B e^{-b S},
\end{eqnarray}
where $A$ and $B$ are constants and the superpotential arises non-perturbatively and describes (possibly many) condensing gauge groups.

At asymptotically large $S$, the effective potential is dominated by a single gaugino condensate superpotential together with the tree-level K\"ahler potential, implying positivity of the scalar potential in the $S \to \infty$ limit (as all other terms are subdominant).
In this case, the positivity of the potential at large $S$ again implies that the presence of any supersymmetric solution at small $S$
would require a turnover of the potential and the existence of a de Sitter critical point.

Finally, the existence of de Sitter critical points does not - of course - imply the existence of de Sitter vacua. Everything that has been said here would be consistent with weaker versions of Eq. (\ref{claim}), which either exclude de Sitter minima, constrain the Hessian of the potential or jointly constrain the first and second derivatives of the potential. Recent examples of these have been discussed in \cite{180610999, 180705193}. Such constraints would be important if they could be shown to be generic properties of string vacua.

\subsection*{Conclusions}\
The claim that string theory does not admit de Sitter critical points is extremely strong.
This note has pointed out that this claim is even stronger than it first appears,
as in cases of interest (including some of the most studied areas of moduli stabilisation, such as IIB flux compactifications) it is equivalent to the absence of supersymmetric AdS solutions. If this conjecture is not to be simply false (or indeed already falsified), the physics that renders such supersymmetric AdS solutions forbidden must indeed be profound.
\subsection*{Acknowledgments}
I thank I\~naki Garc\'{i}a-Etxebarria, Hirosi Ooguri, Eran Palti, Fernando Quevedo, Sav Sethi and Cumrun Vafa for discussions and comments.
\bibliography{SwamplandRefs} 
\end{document}